\documentclass[aps,prl,superscriptaddress,graphicx,amsmath,amssymb,reprint,showkeys]{revtex4-2}

\usepackage[breaklinks=true,colorlinks=true,linkcolor=blue,urlcolor=blue,citecolor=blue]{hyperref}
\usepackage{graphicx}
\usepackage{dcolumn}
\usepackage{bm}

\usepackage[utf8]{inputenc}
\usepackage[T1]{fontenc}
\usepackage{mathptmx}
\usepackage{etoolbox}

\usepackage{epstopdf}
\usepackage[]{units}
\usepackage{upgreek}
\usepackage{csquotes}

\makeatletter
\def\@email#1#2{%
	\endgroup
	\patchcmd{\titleblock@produce}
	{\frontmatter@RRAPformat}
	{\frontmatter@RRAPformat{\produce@RRAP{*#1\href{mailto:#2}{#2}}}\frontmatter@RRAPformat}
	{}{}
}%
\makeatother

\makeatletter
\let\origsection\section
\renewcommand\section{\@ifstar{\starsection}{\nostarsection}}
\newcommand\starsection[1]
{\sectionprelude\origsection*{#1}\sectionpostlude}
\newcommand\sectionprelude{%
	\vspace{0em}}
\newcommand\sectionpostlude{%
	\vspace{-1em}}
\makeatother

\newcommand{\YIG}{yttrium iron garnet}
\newcommand{\SHE}{spin Hall effect}

\newcommand{\STT}{spin-transfer torque}
\newcommand{\SSE}{spin Seebeck effect}

\newcommand{\Gilbert}{Gilbert damping parameter}

\newcommand{\BEC}{Bose-Einstein condensate}
\newcommand{\BLS}{Brillouin-light-scattering}

\newcommand{\uproman}[1]{\uppercase\expandafter{\romannumeral#1}}

\begin{document}

\title{Stimulated amplification of propagating spin waves} 

\author{D. Breitbach}
\email{dbreitba@rptu.de}
\author{M. Schneider}%
\author{B. Heinz}
\author{F. Kohl}
\author{J. Maskill}
\author{L. Scheuer}
\affiliation{Fachbereich Physik and Landesforschungszentrum OPTIMAS, \\Rheinland-Pfälzische Technische Universit{\"a}t Kaiserslautern-Landau, D-67663 Kaiserslautern, Germany}

\author{R. O. Serha}
\affiliation{Fachbereich Physik and Landesforschungszentrum OPTIMAS, \\Rheinland-Pfälzische Technische Universit{\"a}t Kaiserslautern-Landau, D-67663 Kaiserslautern, Germany}
\affiliation{Faculty of Physics, University of Vienna, A-1090 Vienna, Austria}

\author{\mbox{T. Brächer}}
\author{\mbox{B. Lägel}}
\affiliation{Fachbereich Physik and Landesforschungszentrum OPTIMAS, \\Rheinland-Pfälzische Technische Universit{\"a}t Kaiserslautern-Landau, D-67663 Kaiserslautern, Germany}	

\author{C. Dubs}
\affiliation{INNOVENT e.V. Technologieentwicklung, D-07745 Jena, Germany}

\author{V. S. Tiberkevich}
\author{A. N. Slavin}
\affiliation{Department of Physics, Oakland University, Rochester, MI 48309, USA}

\author{A. A. Serga}
\author{B. Hillebrands}
\affiliation{Fachbereich Physik and Landesforschungszentrum OPTIMAS, \\Rheinland-Pfälzische Technische Universit{\"a}t Kaiserslautern-Landau, D-67663 Kaiserslautern, Germany}

\author{A. V. Chumak}
\affiliation{Faculty of Physics, University of Vienna, A-1090 Vienna, Austria}

\author{\mbox{P. Pirro}}
\affiliation{Fachbereich Physik and Landesforschungszentrum OPTIMAS, \\Rheinland-Pfälzische Technische Universit{\"a}t Kaiserslautern-Landau, D-67663 Kaiserslautern, Germany}

\begin{abstract}
Spin-wave amplification techniques are key to the realization of magnon-based computing concepts. We introduce a novel mechanism to amplify spin waves in magnonic nanostructures. Using the technique of rapid cooling, we create a non-equilibrium state in excess of high-energy magnons and demonstrate the stimulated amplification of an externally seeded, propagating spin wave. Using an extended kinetic model, we qualitatively show that the amplification is mediated by an effective energy flux of high energy magnons into the low energy propagating mode, driven by a non-equilibrium magnon distribution.
\end{abstract}

\maketitle 

In recent years, the search for alternative computing methods has sparked interest in analog computing schemes that take advantage of the speed, parallelization and non-binary nature of physical systems \cite{Markovic.2020, Hughes.2019, Torrejon.2017, Shastri.2021}. Spin waves offer unique properties, such as intrinsic nonlinearity, interference, and scalability, making them an efficient carrier of information for such tasks \cite{csaba.2017, Pirro.2021,Chumak.2022}. A variety of spin-wave based devices have been presented, forming the building blocks for more complex magnonic systems \cite{Lenk.2011, Kruglyak.2010, Chumak.2015,Pirro.2021,Chumak.2022}. However, there is a major limitation when combining these components to achieve more intricate functionalities. The spin-wave decay through relaxation mechanisms sets a natural limit to the computational complexity that can be achieved. Therefore, methods are required to efficiently and selectively amplify and sustain spin-wave signals beyond their intrinsic lifetime, enabling for the cascadability of logic elements and achieving a fan-out capability.\newline
Currently, two mechanisms are being explored for spin-wave amplification. The first method is based on the splitting of microwave photons into magnon pairs at half the microwave frequency, known as parametric amplification \cite{Melkov.2001,Bracher.2014,Bracher.2017,Mohseni.2020,Deka.2022}. This technique has been successfully employed to amplify propagating spin waves in macroscopic \cite{Melkov.2001} and microscopic  \cite{Bracher.2014,Bracher.2017,Mohseni.2020} structures. While its fundamental applicability in nanoscopic systems has been demonstrated recently, parametric amplification and its efficiency in these systems is still subject of current research \cite{Heinz.2022}. The second method known to amplify spin waves is based on the injection of spin-currents and the associated \STT{}s (STT). These currents can be generated, for instance, by using an additional magnetic \textquoteleft polarizer\textquoteright\:to spin-polarize a charge current \cite{Berger.1996,Slonczewski.1996}, or by taking advantage of effects such as the \SHE{} (SHE) \cite{Hernandez.2011,Hamadeh.2014,Evelt.2016} or the \SSE{} (SSE) \cite{Uchida.2008, Hernandez.2012}. This method allows for partial or even overcompensation of the spin-wave damping \cite{Merbouche.2023}. A drawback of using the SHE for this approach is the requirement of spin injection layers, which naturally reduce the spin-wave lifetime through spin-pumping and microwave eddy currents, thereby counteracting the desired amplification.\newline
In this study, we take a fundamentally different approach to spin-wave amplification which is based on the stimulated emission of magnons. In thermodynamic equilibrium, typically, a stimulated amplification process cannot achieve a net energy gain. Instead, analogous to the amplification in a LASER, a non-equilibrium energy distribution is required. To achieve such state, we use the method of rapid cooling \cite{RCBEC, Schneider.2021, Schneider.2021b}: A microscopic \YIG{} (YIG) waveguide is locally heated by a short DC current pulse in an adjacent metal. When the DC pulse is turned off, the heat rapidly dissipates to the surrounding material, causing a sudden decrease in lattice temperature. This process is approximately instant compared to the magnon lifetime in YIG, causing an imbalance between the lattice and the magnon system. In the process of the reduction of the free energy of the system, the excess magnons are redistributed towards the bottom of the magnon spectrum. This process can trigger the formation of a magnon \BEC{} (BEC) \cite{RCBEC, Schneider.2021b}, or stabilize a nonlinear, localized magnonic bullet mode below the bottom of the magnon spectrum \cite{Schneider.2021}. We have shown that these phenomena can be understood in the simplified picture of a quasi-equilibrium approximation of the rapid cooling process using a global chemical potential $\mu$. Here, we theoretically take into account the non-equilibrium nature of the rapid cooling process by introducing a local chemical potential $\mu_{\mathrm{\mathbf{k}}}$ and local temperature $T_{\mathrm{\mathbf{k}}}$ to the kinetic model developed in Ref. \cite{RCBEC}. We find that a net energy flow to spin-wave modes even above the minimum frequency of the magnon spectrum can occur, driven by the non-equilibrium excess of high-energy magnons. Experimentally, we investigate this process by seeding the rapidly cooled magnon gas with an externally excited, propagating spin-wave packet. This serves as an attractor state to the redistribution process, leading to stimulated amplification.\newline
Previous approaches to spin-wave lasing have utilized electron-mediated mechanisms such as spin-wave/carrier-wave interactions \cite{Robinson.1970, Nunes.1982, Souto.2001}, spin-transfer torques \cite{Berger.1996, Collet.2016, Tsoi.2000, Doornenbal.2019} or an effect known as quantum amplification \cite{Danilov.1980, Danilov.2002} to pump the spin-wave system. The most crucial difference between our experiments and previous approaches is that we realize a thermodynamic LASER where the energy is provided by an incoherently and thermally pumped, non-equilibrium magnon distribution.\newline
\begin{figure}
	\includegraphics{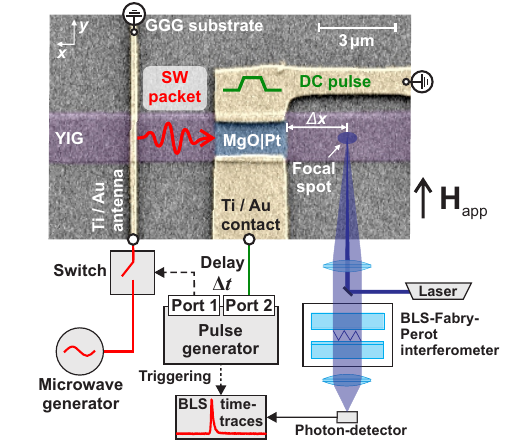}%
	\caption{\label{fig:Structure} Colorized SEM micrograph of a YIG waveguide structure (purple) similar to the one under study. The MgO|Pt amplifier pad (blue) is placed on top of the waveguide and connected to a pulse generator, allowing for the application of short DC pulses. A stripline antenna is used for spin-wave excitation by RF pulses.}%
\end{figure}
The investigated microstructure and the experimental setup are shown in Fig. \ref{fig:Structure}. The structure consists of a 34-nm-thick and 2-$\upmu$m-wide \YIG{} (YIG) waveguide, fabricated by argon ion milling (see, for instance, Ref. \cite{Heinz.2020}) from an LPE-grown YIG film  \cite{Dubs.2020}. Standard vector-network-analyzer ferromagnetic resonance measurements were performed on a reference pad on the sample, yielding a \Gilbert{} of $\alpha = 1.8 \times 10^{-4}$. The waveguide is magnetized transversely by an external field of $\upmu_{\mathrm{0}}\mathbf{H}_{\mathrm{app}} = \unit[188]{mT}$. A 2-$\upmu$m-long area of the waveguide is covered by a pad of MgO|Pt (5|\unit[7]{nm}) grown by molecular beam epitaxy, which in the following will be referred to as the \textit{amplifier pad}. Electrical contacts to the amplifier pad and a stripline antenna were structured from a Ti|Au material system.\newline
In the experiment, short DC heating pulses of length \mbox{$\tau_{\mathrm{DC}} = \unit[15]{ns}$} and voltages of up to $U_{\mathrm{DC}}=\unit[1.1]{V}$, corresponding to a current density of $j_{\mathrm{DC}} = \unit[(1.56\pm0.3)\times 10^{12}]{A m^{-2}}$, are used to locally increase the temperature by Joule heating, see Supplemental Material \cite{Supplemental}. The MgO interlayer serves as a spin-diffusion barrier, preventing spin injection from the Pt layer into the waveguide \cite{Mosendz.2010,Ruiz-Calaforra.2015}. Subsequently, after the heating pulse is turned off, a rapid cooling process takes place until a thermodynamic equilibrium is achieved. A microwave switch is used to generate a short microwave pulse ($\tau_{\mathrm{RF}} \approx \unit[10]{ns}$) with a carrier frequency above the bottom of the spin-wave spectrum. In the following experiments, this frequency is $f_{\mathrm{MW}} = \unit[7.225]{GHz}$, see Supplemental Material \cite{Supplemental} for other excitation frequencies. This pulse then excites a packet of propagating spin waves in the magnonic waveguide below the RF antenna.\newline
The resulting spin dynamics is studied with time- and space-resolved \BLS{} (BLS) spectroscopy, using a laser beam of $\lambda = \unit[457]{nm}$ and $P = \unit[(5.0\pm0.5)]{mW}$. The employed BLS setup hosts an upside-down sample geometry, allowing to focus the laser (spot size \unit[400]{nm}) from the backside of the sample through the transparent gadolinium gallium garnet (GGG) substrate onto the YIG waveguide \cite{RCBEC, Schneider.2021, Schneider.2021b, Heinz.2020,QWang.2019}. This enables measurements in the region covered by the amplifier pad.

\begin{figure}[t]
	\includegraphics{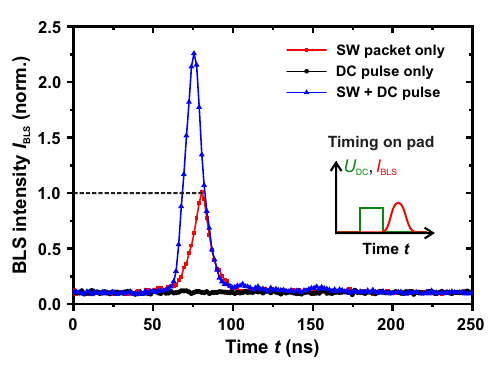}%
	\caption{\label{fig:Motivation_amp} BLS intensity as a function of time, normalized to the maximum of the red curve. The measurement was performed $\Delta x =\unit[1.7]{\upmu m}$ behind the amplifier pad using three different pulse configurations. In the case of 'SW + DC pulse', the spin-wave (SW) packet was timed to arrive in the amplifier region just after the DC heating pulse. Parameters: $P_{\mathrm{RF}} = \unit[-20]{dBm}$, $U_{\mathrm{DC}} = \unit[1.1]{V}$.}
\end{figure}
After the spin-wave packet is excited, it propagates along the waveguide and passes the amplifier pad. Three different pulse configurations are compared: the application of the RF pulse only, the heating pulse only, or both together. The resulting time-dependent BLS intensity, measured $\Delta x =\unit[1.7]{\upmu m}$ behind the pad, is shown in Fig. \ref{fig:Motivation_amp}. For the case of only the RF pulse applied (red curve), we observe a spin-wave packet of duration $\tau_{\mathrm{FWHM}}=\unit[15]{ns}$, to which the intensity scale is normalized. When additionally a heating pulse is sent through the amplifier pad, the result changes quite drastically. When timed such that the spin-wave packet is below the amplifier pad just when the heating pulse ends, i.e. in the moment of the largest cooling rate, the spin-wave packet is significantly amplified. The resulting BLS intensity is more than twice as high compared to the case without a heating pulse (blue curve). However, there is no spin-wave emission detected when only the heating pulse is applied. This indicates that the observed amplification is caused by an interaction between the spin-wave packet and the consequences of the heating pulse.\newline
\begin{figure}[t]
	\includegraphics{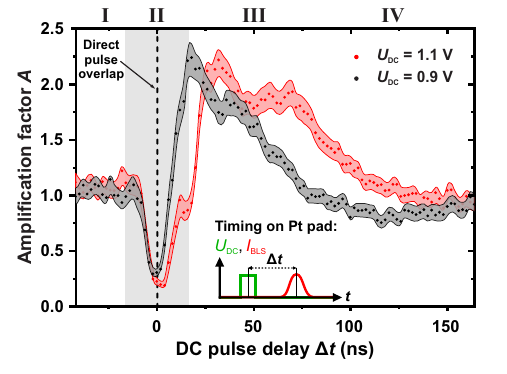}%
	\caption{\label{fig:Delay} Amplification factor $A$ as a function of the time delay $\Delta t$ between spin-wave packet and the DC heating pulse using a continuous representation of error bars. The delay $\Delta t$ is set to be zero for minimal transmission which marks direct pulse overlap on the amplifier pad, the shaded area marks the uncertainty regime with full or partial pulse overlap. Parameters: $\Delta x =\unit[1.7]{\upmu m}$, $P_{\mathrm{RF}} = \unit[-20]{dBm}$.}
\end{figure}
We quantitatively investigate this amplification as a function of the time delay $\Delta t$ between the RF and heating pulse, see Fig. \ref{fig:Delay}, using the amplification factor $A$, defined as $A = {I_{\mathrm{BLS}}(U_{\mathrm{DC}})}/{I_{\mathrm{BLS}}(U_{\mathrm{DC}}=\unit[0]{V})}$. As the spin-wave packet takes time to propagate through the waveguide, there is an extended window of overlap with the DC heating pulse on the amplifier pad. For the sake of simplicity, the RF pulse delay $\Delta t$ is set to be zero for minimal transmission. In the resulting graph, four different regimes can be observed.\newline
\textbf{Regime \uproman{1}}
The spin-wave packet passes the amplifier pad before the heating pulse is applied. As a result, no effect on the amplification factor is observed.\newline
\textbf{Regime \uproman{2}}
The spin-wave packet directly overlaps with the heating pulse below the amplifier pad. In this regime, it is affected by the dynamic temperature increase during the heating pulse, and the measured BLS intensity behind the pad is drastically decreased. The total overlap time is given by the length of the heating pulse and the spin-wave packet and by the time it takes the spin-wave packet to pass the amplifier pad. For $\Delta t > 0$, partial overlap with the rapid cooling process occurs, and the transition to the next regime begins.\newline
\textbf{Regime \uproman{3}}
In this regime, the amplification effect is observed. It is strongest when the spin-wave packet reaches the amplifier pad just after the end of the heating pulse and decreases for larger $\Delta t$ as the cooling rate decreases. Note that this window of amplification is much longer than the actual heating pulse.\newline
\textbf{Regime \uproman{4}}
At the end of regime \uproman{3}, the amplification vanishes, and $A$ drops slightly below unity, before recovering to its value of \mbox{regime \uproman{1}}.\newline
These results show the presence of two effects. On the one hand, the direct influence of the DC heating pulse overlapping with the spin-wave packet is evident, leading to a significant decrease in spin-wave transmission in regime \uproman{2}. This is caused by the forced frequency downshift of the spin-wave packet on the amplifier pad during the time-dependent temperature increase induced by the DC heating pulse, and the consequent decrease in saturation magnetization $M_{\mathrm{S}}$ and exchange stiffness $D_{\mathrm{ex}}$ of YIG \cite{SAGA}, see Supplemental Material \cite{Supplemental}. On the other hand, there is the amplification effect, which is observed to be maximal during the rapid cooling phase in regime \uproman{3}, indicating that the amplification is related to the cooling rate of the system. Hence, the effect arises when the spin-wave packet propagates below the amplifier pad after the heating pulse has ended, proving that the effect is of purely thermal origin. Other effects that could be involved, such as the SHE, the SSE, or \O ersted fields related to the DC heating pulse, would be maximized during direct pulse overlap and therefore can be ruled out from contributing to the amplification process. This experiment shows that the amplitude of the spin-wave signal can be controlled by a factor in a range between 0.2 (attenuation) and almost 2.5 (amplification) simply by adjusting the relative timing to the heating pulse. \newline
\begin{figure}[t]
	\includegraphics{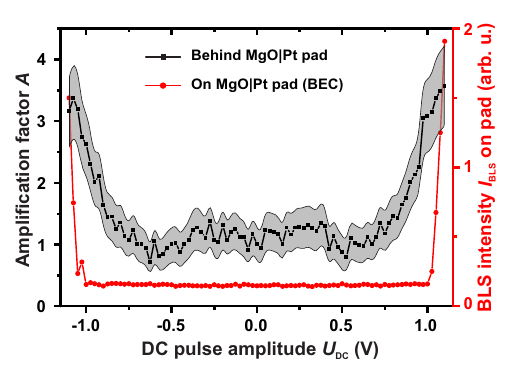}%
	\caption{\label{fig:DC} \textbf{Black curve, left scale}: Amplification factor $A$ as a function of the applied DC heating pulse amplitude $U_{\mathrm{DC}}$. Measurement performed for $\Delta t = \unit[50]{ns}$, $P_{\mathrm{RF}} = \unit[-27]{dBm}$ and in a distance of $\Delta x =\unit[3]{\upmu m}$ to the amplifier pad - slightly more than the previous measurements, see Supplemental Material \cite{Supplemental} for a discussion on the impact on the amplification factor. \textbf{Red curve, right scale}: BLS intensity measured on the amplifier pad using only a DC heating pulse but no spin-wave excitation.}%
\end{figure}
The heating power scales approximately quadratically with $U_{\mathrm{DC}}$, the influence of which is seen when comparing $U_{\mathrm{DC}}=\unit[1.1]{V}$ to $U_{\mathrm{DC}}=\unit[0.9]{V}$ in Fig. \ref{fig:Delay}. The temperatures reached for $U_{\mathrm{DC}} = \unit[1.1]{V}$ are higher, leading to a more pronounced attenuation of the spin-wave packet in regime \uproman{2} and the cooling phase is longer as more heat needs to be dissipated, leading to an extended amplification phase \uproman{3}. In this phase, the formation of a second amplification plateau can be observed. This could be a product of indirect magnon redistribution processes, as it is correlated to the formation of a \BEC{} in the region below the amplifier pad above the threshold value of $|U^{\mathrm{BEC}}_{\mathrm{Th}}| = \unit[1.0]{V}$ \cite{RCBEC, Schneider.2021}. The BEC formation threshold is apparent from a measurement of the BLS intensity in the amplifier pad region itself, which is shown in Fig. \ref{fig:DC} by the red curve. Additionally, the amplification factor is shown, measured as a function of $U_{\mathrm{DC}}$ and using a fixed delay of $\Delta t = \unit[50]{ns}$. For small amplitudes $|U_{\mathrm{DC}}|<\unit[0.4]{V}$, the heating pulse does not affect the intensity of the spin-wave packet. For intermediate amplitudes, $\unit[0.4]{V}<|U_{\mathrm{DC}}|<\unit[0.6]{V}$, a slight attenuation is observed, which could be due to residual heat and the consequent impact on the magnetic properties of the waveguide. For amplitudes $|U_{\mathrm{DC}}|>\unit[0.6]{V}$, the amplification process dominates, which is symmetric with respect to the sign of $U_{\mathrm{DC}}$. This again highlights the thermal origin of the amplification effect. In contrast to the BEC formation \cite{RCBEC}, the amplification effect does not have a clear threshold value and scales with the applied heating power.\newline

In the following, an extension of the kinetic model developed in Ref. \cite{RCBEC} is derived for the simplified case of a thin film, which accounts for the non-equilibrium nature of the rapid cooling process. The interaction of a propagating spin-wave mode with wavevector $\mathbf{k}$ with a rapidly cooled magnon gas can be qualitatively analyzed using the kinetic equations for the population of magnon states \cite{Lvov.1994}. The influence of 4-magnon scattering leads to an additional energy flux $G_{\mathrm{k}}$ for the magnon population $\tilde{n}_{\mathrm{k}}$:
\begin{equation}\label{Equ:kinetic}
\mathrm{d}\tilde{n}_{\mathrm{k}}/\mathrm{d}t = -2(\Gamma_{\mathrm{k}}+G_{\mathrm{k}})\tilde{n}_{\mathrm{k}},
\end{equation}
see Supplemental Material \cite{Supplemental} for details. Here, $\Gamma_{\mathrm{k}}$ is the spin-lattice relaxation rate. If the magnon gas is in thermodynamic equilibrium, $G_{\mathrm{k}}$ can be written as
\begin{multline}\label{Equ:Gk123}
G_{\mathrm{k}} = \frac{1}{2}\sum_{123}T_{\mathrm{k123}}n_{\mathrm{1}}n_{\mathrm{2}}n_{\mathrm{3}}\\ \times \mathrm{exp}\left(\frac{\hbar\omega_{\mathrm{1}}-\mu}{k_{\mathrm{B}}T}\right)\left[\mathrm{exp}\left(\frac{\hbar\omega_{\mathrm{k}}-\mu}{k_{\mathrm{B}}T}\right)-1\right]
\end{multline}
where $\mathbf{\mathrm{1}}\equiv\mathbf{k}_{\mathrm{1}}$, etc., where the summation includes all scattering processes satisfying energy $\omega_{\mathrm{k}}+\omega_{\mathrm{1}} = \omega_{\mathrm{2}} + \omega_{\mathrm{3}}$ and momentum $\mathbf{k} + \mathbf{k}_{\mathrm{1}} = \mathbf{k}_{\mathrm{2}}+\mathbf{k}_{\mathrm{3}}$ conservation laws. $T_{\mathrm{k123}}>0$ are the 4-magnon scattering amplitudes, $n_{\mathrm{j}}$ is the population of state $\mathbf{k}_{\mathrm{j}}$ in the magnon gas, and $\mu$ and $T$ are the chemical potential and temperature of the gas, respectively.\newline
In the approximation of thermodynamic equilibrium, the chemical potential cannot exceed the gap in the magnon spectrum, meaning that $\hbar\omega_{\mathrm{k}}>\mu$ and $G_{\mathrm{k}}>0$, thus the magnon-magnon interaction leads to an energy loss of the propagating spin-wave mode $\tilde{n}_{\mathrm{k}}$. In the case of a non-equilibrium magnon distribution with excess of high energy magnons, however, this situation can change, as an energy flux from high energy magnon modes into the propagating spin-wave mode $\mathbf{k}$ can arise. Here, we theoretically take into account the non-equilibrium nature of the rapid cooling process by introducing a local chemical potential $\mu_{\mathrm{\mathbf{k}}}$ and local temperature $T_{\mathrm{\mathbf{k}}}$ to the kinetic model developed in Ref. \cite{RCBEC} and derive an equation for $G_{\mathrm{\mathbf{k}}}$ similar to Eq. (\ref{Equ:Gk123}), see Supplemental Material \cite{Supplemental} for details. Our simulations of the population dynamics in the rapid cooling process show that, while the effective temperature $T_{\mathrm{\mathbf{k}}}$ remains almost uniform across the magnon spectrum, see Supplemental Material \cite{Supplemental}, the local chemical potential $\mu_{\mathrm{\mathbf{k}}}$ develops a significant non-uniformity.
\begin{figure}[t]
	\includegraphics[scale=0.66]{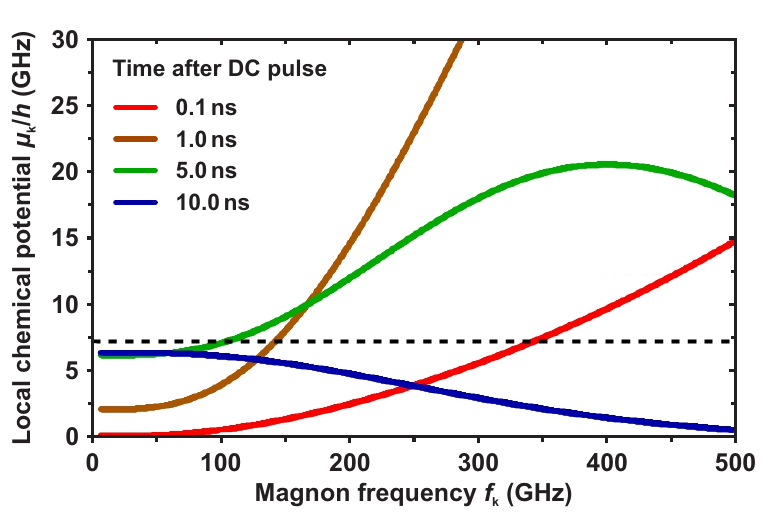}%
	\caption{\label{fig:ChemPot} Dependence of the effective local chemical potential $\mu_{\mathrm{\mathbf{k}}}$ on the magnon frequency $f_{\mathrm{\mathbf{k}}}$ for different delay times after the onset of the rapid cooling process. Dashed horizontal line shows the minimum frequency of the magnon spectrum $f_{\mathrm{min}} = \unit[7.20]{GHz}$.}
\end{figure}\newline
Figure \ref{fig:ChemPot} shows an example dependence of $\mu_{\mathrm{\mathbf{k}}}$ on the magnon frequency $\omega_{\mathrm{\mathbf{k}}}$ for different delay times after the start of the rapid cooling process. The local chemical potential $\mu_{\mathrm{\mathbf{k}}}$ at low-energies, i.e. in the BLS observable frequency range, increases up to the magnon gap $\hbar\omega_{\mathrm{min}}$ but never exceeds it, which is consistent with Ref. \cite{RCBEC}. The chemical potential of higher-energy magnon states, however, significantly exceeds $\hbar\omega_{\mathrm{min}}$ at intermediate times $t\sim\unit[1-10]{ns}$ after the onset of the rapid cooling process, which indicates the non-equilibrium character of the process. As discussed in detail in the Supplemental Material \cite{Supplemental}, this imbalance of $\mu_{\mathrm{\mathbf{k}}}$ opens new channels for an effective energy flux from a high energy mode to the propagating mode $\tilde{n}_{\mathrm{k}}$, e.g. in processes such as $\omega_{\mathrm{2}}+\omega_{\mathrm{2}} \leftrightarrow \omega_{\mathrm{k}} + \omega_{\mathrm{1}}$, where two degenerate magnons $\omega_{\mathrm{2}}$ lead to the population of $\omega_{\mathrm{k}}$ and an idler wave $\omega_{\mathrm{1}}$. Respectively, the terms in Eq. (\ref{Equ:Gk123}) describing scattering with these higher-energy magnon states become negative, which can lead to the sign reversal of the magnon-magnon relaxation rate: $G_{\mathrm{k123}}<0$. Following Eq. (\ref{Equ:kinetic}), the high magnon population of the excited, propagating mode then acts as an attractor state to these scattered magnons.\newline
This \textquoteleft overshoot\textquoteright\:of the local chemical potential $\mu_{k}>\hbar\omega_{min}$ at higher energy magnon states and the corresponding sign reversal of the magnon-magnon relaxation term qualitatively explains the observed amplification of propagating spin waves during the rapid cooling process. The attractive nature of the process implies that the frequency of the amplified mode is set by the incident spin-wave packet. Experimentally, the frequency of the spin-wave packet is found to be preserved in the amplification process, and the amplification factor is a smooth function of the excited spin-wave frequency, see Supplemental Material \cite{Supplemental}.\newline
In conclusion, our study has demonstrated the stimulated amplification of propagating spin waves using the technique of rapid cooling. The mechanism is based on an interaction between a propagating spin-wave mode and the rapidly cooled magnon gas. Describing the non-equilibrium population dynamics in an extended kinetic model with a local chemical potential $\mu_{\mathrm{k}}$, we have qualitatively shown that this interaction results in a net energy flux driven by the scattering of excess high energy magnons. Importantly, our study has shown that the effect is achieved by a purely thermal stimulus and can be realized using spin-pumping free amplifier materials. Our results provide a proof of principle for a magnonic stimulated amplification process analogous to the LASER, pointing towards a new class of magnonic amplifiers with a potential for high efficiency and practicality.\newline

\begin{acknowledgements}
This research was funded by the European Research Council within the Starting Grant No. 101042439 "CoSpiN" and No. 678309 “MagnonCircuits”, by the Deutsche Forschungsgemeinschaft (DFG, German Research Foundation) within the Transregional Collaborative Research Center—TRR 173–268565370 “Spin + X” (project B01 and B04) and the project 271741898 and by the Austrian Science Fund (FWF) through Project No. I 4696-N. The authors acknowledge support by the Max Planck Graduate Center with the Johannes Gutenberg-Universität Mainz (MPGC). This work was supported in part by the Air Force Office of Scientific Research under the MURI grant \# FA9550-19-1-0307.
\end{acknowledgements}

\section*{References}

\begin{thebibliography}{43}%
	\makeatletter
	\providecommand \@ifxundefined [1]{%
		\@ifx{#1\undefined}
	}%
	\providecommand \@ifnum [1]{%
		\ifnum #1\expandafter \@firstoftwo
		\else \expandafter \@secondoftwo
		\fi
	}%
	\providecommand \@ifx [1]{%
		\ifx #1\expandafter \@firstoftwo
		\else \expandafter \@secondoftwo
		\fi
	}%
	\providecommand \natexlab [1]{#1}%
	\providecommand \enquote  [1]{``#1''}%
	\providecommand \bibnamefont  [1]{#1}%
	\providecommand \bibfnamefont [1]{#1}%
	\providecommand \citenamefont [1]{#1}%
	\providecommand \href@noop [0]{\@secondoftwo}%
	\providecommand \href [0]{\begingroup \@sanitize@url \@href}%
	\providecommand \@href[1]{\@@startlink{#1}\@@href}%
	\providecommand \@@href[1]{\endgroup#1\@@endlink}%
	\providecommand \@sanitize@url [0]{\catcode `\\12\catcode `\$12\catcode
		`\&12\catcode `\#12\catcode `\^12\catcode `\_12\catcode `\%12\relax}%
	\providecommand \@@startlink[1]{}%
	\providecommand \@@endlink[0]{}%
	\providecommand \url  [0]{\begingroup\@sanitize@url \@url }%
	\providecommand \@url [1]{\endgroup\@href {#1}{\urlprefix }}%
	\providecommand \urlprefix  [0]{URL }%
	\providecommand \Eprint [0]{\href }%
	\providecommand \doibase [0]{https://doi.org/}%
	\providecommand \selectlanguage [0]{\@gobble}%
	\providecommand \bibinfo  [0]{\@secondoftwo}%
	\providecommand \bibfield  [0]{\@secondoftwo}%
	\providecommand \translation [1]{[#1]}%
	\providecommand \BibitemOpen [0]{}%
	\providecommand \bibitemStop [0]{}%
	\providecommand \bibitemNoStop [0]{.\EOS\space}%
	\providecommand \EOS [0]{\spacefactor3000\relax}%
	\providecommand \BibitemShut  [1]{\csname bibitem#1\endcsname}%
	\let\auto@bib@innerbib\@empty
	\bibitem [{\citenamefont {Markovi{\'{c}}}\ \emph {et~al.}(2020)\citenamefont
		{Markovi{\'{c}}}, \citenamefont {Mizrahi}, \citenamefont {Querlioz},\ and\
		\citenamefont {Grollier}}]{Markovic.2020}%
	\BibitemOpen
	\bibfield  {author} {\bibinfo {author} {\bibfnamefont {D.}~\bibnamefont
			{Markovi{\'{c}}}}, \bibinfo {author} {\bibfnamefont {A.}~\bibnamefont
			{Mizrahi}}, \bibinfo {author} {\bibfnamefont {D.}~\bibnamefont {Querlioz}},\
		and\ \bibinfo {author} {\bibfnamefont {J.}~\bibnamefont {Grollier}},\
	}\bibfield  {title} {\bibinfo {title} {{Physics for neuromorphic
				computing}},\ }\href {https://doi.org/10.1038/s42254-020-0208-2} {\bibfield
		{journal} {\bibinfo  {journal} {Nat. Rev. Phys.}\ }\textbf {\bibinfo {volume}
			{2}},\ \bibinfo {pages} {499} (\bibinfo {year} {2020})}\BibitemShut {NoStop}%
	\bibitem [{\citenamefont {Hughes}\ \emph {et~al.}(2019)\citenamefont {Hughes},
		\citenamefont {Williamson}, \citenamefont {Minkov},\ and\ \citenamefont
		{Fan}}]{Hughes.2019}%
	\BibitemOpen
	\bibfield  {author} {\bibinfo {author} {\bibfnamefont {T.~W.}\ \bibnamefont
			{Hughes}}, \bibinfo {author} {\bibfnamefont {I.~A.~D.}\ \bibnamefont
			{Williamson}}, \bibinfo {author} {\bibfnamefont {M.}~\bibnamefont {Minkov}},\
		and\ \bibinfo {author} {\bibfnamefont {S.}~\bibnamefont {Fan}},\ }\bibfield
	{title} {\bibinfo {title} {{Wave physics as an analog recurrent neural
				network}},\ }\href {https://doi.org/10.1126/sciadv.aay6946} {\bibfield
		{journal} {\bibinfo  {journal} {Sci. Adv.}\ }\textbf {\bibinfo {volume}
			{5}},\ \bibinfo {pages} {eaay6946} (\bibinfo {year} {2019})}\BibitemShut
	{NoStop}%
	\bibitem [{\citenamefont {Torrejon}\ \emph {et~al.}(2017)\citenamefont
		{Torrejon}, \citenamefont {Riou}, \citenamefont {Araujo}, \citenamefont
		{Tsunegi}, \citenamefont {Khalsa}, \citenamefont {Querlioz}, \citenamefont
		{Bortolotti}, \citenamefont {Cros}, \citenamefont {Yakushiji}, \citenamefont
		{Fukushima}, \citenamefont {Kubota}, \citenamefont {Yuasa}, \citenamefont
		{Stiles},\ and\ \citenamefont {Grollier}}]{Torrejon.2017}%
	\BibitemOpen
	\bibfield  {author} {\bibinfo {author} {\bibfnamefont {J.}~\bibnamefont
			{Torrejon}}, \bibinfo {author} {\bibfnamefont {M.}~\bibnamefont {Riou}},
		\bibinfo {author} {\bibfnamefont {F.~A.}\ \bibnamefont {Araujo}}, \bibinfo
		{author} {\bibfnamefont {S.}~\bibnamefont {Tsunegi}}, \bibinfo {author}
		{\bibfnamefont {G.}~\bibnamefont {Khalsa}}, \bibinfo {author} {\bibfnamefont
			{D.}~\bibnamefont {Querlioz}}, \bibinfo {author} {\bibfnamefont
			{P.}~\bibnamefont {Bortolotti}}, \bibinfo {author} {\bibfnamefont
			{V.}~\bibnamefont {Cros}}, \bibinfo {author} {\bibfnamefont {K.}~\bibnamefont
			{Yakushiji}}, \bibinfo {author} {\bibfnamefont {A.}~\bibnamefont
			{Fukushima}}, \bibinfo {author} {\bibfnamefont {H.}~\bibnamefont {Kubota}},
		\bibinfo {author} {\bibfnamefont {S.}~\bibnamefont {Yuasa}}, \bibinfo
		{author} {\bibfnamefont {M.~D.}\ \bibnamefont {Stiles}},\ and\ \bibinfo
		{author} {\bibfnamefont {J.}~\bibnamefont {Grollier}},\ }\bibfield  {title}
	{\bibinfo {title} {{Neuromorphic computing with nanoscale spintronic
				oscillators}},\ }\href {https://doi.org/10.1038/nature23011} {\bibfield
		{journal} {\bibinfo  {journal} {Nature}\ }\textbf {\bibinfo {volume} {547}},\
		\bibinfo {pages} {428} (\bibinfo {year} {2017})}\BibitemShut {NoStop}%
	\bibitem [{\citenamefont {Shastri}\ \emph {et~al.}(2021)\citenamefont
		{Shastri}, \citenamefont {Tait}, \citenamefont {Ferreira~de Lima},
		\citenamefont {Pernice}, \citenamefont {Bhaskaran}, \citenamefont {Wright},\
		and\ \citenamefont {Prucnal}}]{Shastri.2021}%
	\BibitemOpen
	\bibfield  {author} {\bibinfo {author} {\bibfnamefont {B.~J.}\ \bibnamefont
			{Shastri}}, \bibinfo {author} {\bibfnamefont {A.~N.}\ \bibnamefont {Tait}},
		\bibinfo {author} {\bibfnamefont {T.}~\bibnamefont {Ferreira~de Lima}},
		\bibinfo {author} {\bibfnamefont {W.~H.~P.}\ \bibnamefont {Pernice}},
		\bibinfo {author} {\bibfnamefont {H.}~\bibnamefont {Bhaskaran}}, \bibinfo
		{author} {\bibfnamefont {C.~D.}\ \bibnamefont {Wright}},\ and\ \bibinfo
		{author} {\bibfnamefont {P.~R.}\ \bibnamefont {Prucnal}},\ }\bibfield
	{title} {\bibinfo {title} {{Photonics for artificial intelligence and
				neuromorphic computing}},\ }\href
	{https://doi.org/10.1038/s41566-020-00754-y} {\bibfield  {journal} {\bibinfo
			{journal} {Nat. Photonics}\ }\textbf {\bibinfo {volume} {15}},\ \bibinfo
		{pages} {102} (\bibinfo {year} {2021})}\BibitemShut {NoStop}%
	\bibitem [{\citenamefont {Csaba}\ \emph {et~al.}(2017)\citenamefont {Csaba},
		\citenamefont {Papp},\ and\ \citenamefont {Porod}}]{csaba.2017}%
	\BibitemOpen
	\bibfield  {author} {\bibinfo {author} {\bibfnamefont {G.}~\bibnamefont
			{Csaba}}, \bibinfo {author} {\bibfnamefont {A.}~\bibnamefont {Papp}},\ and\
		\bibinfo {author} {\bibfnamefont {W.}~\bibnamefont {Porod}},\ }\bibfield
	{title} {\bibinfo {title} {{Perspectives of using spin waves for computing
				and signal processing}},\ }\href
	{https://doi.org/10.1016/j.physleta.2017.02.042} {\bibfield  {journal}
		{\bibinfo  {journal} {Phys. Lett. A}\ }\textbf {\bibinfo {volume} {381}},\
		\bibinfo {pages} {1471} (\bibinfo {year} {2017})}\BibitemShut {NoStop}%
	\bibitem [{\citenamefont {Pirro}\ \emph {et~al.}(2021)\citenamefont {Pirro},
		\citenamefont {Vasyuchka}, \citenamefont {Serga},\ and\ \citenamefont
		{Hillebrands}}]{Pirro.2021}%
	\BibitemOpen
	\bibfield  {author} {\bibinfo {author} {\bibfnamefont {P.}~\bibnamefont
			{Pirro}}, \bibinfo {author} {\bibfnamefont {V.}~\bibnamefont {Vasyuchka}},
		\bibinfo {author} {\bibfnamefont {A.~A.}\ \bibnamefont {Serga}},\ and\
		\bibinfo {author} {\bibfnamefont {B.}~\bibnamefont {Hillebrands}},\
	}\bibfield  {title} {\bibinfo {title} {{Advances in coherent magnonics}},\
	}\href {https://doi.org/10.1038/s41578-021-00332-w} {\bibfield  {journal}
		{\bibinfo  {journal} {Nat. Rev. Mater.}\ }\textbf {\bibinfo {volume} {6}},\
		\bibinfo {pages} {1114} (\bibinfo {year} {2021})}\BibitemShut {NoStop}%
	\bibitem [{\citenamefont {Chumak}\ \emph {et~al.}(2022)\citenamefont {Chumak},
		\citenamefont {Kabos}, \citenamefont {Wu} \emph {et~al.}}]{Chumak.2022}%
	\BibitemOpen
	\bibfield  {author} {\bibinfo {author} {\bibfnamefont {A.~V.}\ \bibnamefont
			{Chumak}}, \bibinfo {author} {\bibfnamefont {P.}~\bibnamefont {Kabos}},
		\bibinfo {author} {\bibfnamefont {M.}~\bibnamefont {Wu}}, \emph {et~al.},\
	}\bibfield  {title} {\bibinfo {title} {{Advances in magnetics roadmap on
				spin-wave computing}},\ }\href {https://doi.org/10.1109/TMAG.2022.3149664}
	{\bibfield  {journal} {\bibinfo  {journal} {IEEE Trans. Magn.}\ }\textbf
		{\bibinfo {volume} {58}},\ \bibinfo {pages} {1} (\bibinfo {year}
		{2022})}\BibitemShut {NoStop}%
	\bibitem [{\citenamefont {Lenk}\ \emph {et~al.}(2011)\citenamefont {Lenk},
		\citenamefont {Ulrichs}, \citenamefont {Garbs},\ and\ \citenamefont
		{Münzenberg}}]{Lenk.2011}%
	\BibitemOpen
	\bibfield  {author} {\bibinfo {author} {\bibfnamefont {B.}~\bibnamefont
			{Lenk}}, \bibinfo {author} {\bibfnamefont {H.}~\bibnamefont {Ulrichs}},
		\bibinfo {author} {\bibfnamefont {F.}~\bibnamefont {Garbs}},\ and\ \bibinfo
		{author} {\bibfnamefont {M.}~\bibnamefont {Münzenberg}},\ }\bibfield
	{title} {\bibinfo {title} {{The building blocks of magnonics}},\ }\href
	{https://doi.org/https://doi.org/10.1016/j.physrep.2011.06.003} {\bibfield
		{journal} {\bibinfo  {journal} {Phys. Rep.}\ }\textbf {\bibinfo {volume}
			{507}},\ \bibinfo {pages} {107} (\bibinfo {year} {2011})}\BibitemShut
	{NoStop}%
	\bibitem [{\citenamefont {Kruglyak}\ \emph {et~al.}(2010)\citenamefont
		{Kruglyak}, \citenamefont {Demokritov},\ and\ \citenamefont
		{Grundler}}]{Kruglyak.2010}%
	\BibitemOpen
	\bibfield  {author} {\bibinfo {author} {\bibfnamefont {V.~V.}\ \bibnamefont
			{Kruglyak}}, \bibinfo {author} {\bibfnamefont {S.~O.}\ \bibnamefont
			{Demokritov}},\ and\ \bibinfo {author} {\bibfnamefont {D.}~\bibnamefont
			{Grundler}},\ }\bibfield  {title} {\bibinfo {title} {{Magnonics}},\ }\href
	{https://doi.org/10.1088/0022-3727/43/26/264001} {\bibfield  {journal}
		{\bibinfo  {journal} {J. Phys. D}\ }\textbf {\bibinfo {volume} {43}},\
		\bibinfo {pages} {264001} (\bibinfo {year} {2010})}\BibitemShut {NoStop}%
	\bibitem [{\citenamefont {Chumak}\ \emph {et~al.}(2015)\citenamefont {Chumak},
		\citenamefont {Vasyuchka}, \citenamefont {Serga},\ and\ \citenamefont
		{Hillebrands}}]{Chumak.2015}%
	\BibitemOpen
	\bibfield  {author} {\bibinfo {author} {\bibfnamefont {A.}~\bibnamefont
			{Chumak}}, \bibinfo {author} {\bibfnamefont {V.}~\bibnamefont {Vasyuchka}},
		\bibinfo {author} {\bibfnamefont {A.}~\bibnamefont {Serga}},\ and\ \bibinfo
		{author} {\bibfnamefont {B.}~\bibnamefont {Hillebrands}},\ }\bibfield
	{title} {\bibinfo {title} {{Magnon spintronics}},\ }\href
	{https://doi.org/https://doi.org/10.1038/nphys3347} {\bibfield  {journal}
		{\bibinfo  {journal} {Nat. Phys.}\ }\textbf {\bibinfo {volume} {11}},\
		\bibinfo {pages} {453} (\bibinfo {year} {2015})}\BibitemShut {NoStop}%
	\bibitem [{\citenamefont {Melkov}\ \emph {et~al.}(2001)\citenamefont {Melkov},
		\citenamefont {Serga}, \citenamefont {Tiberkevich}, \citenamefont
		{Kobljanskij},\ and\ \citenamefont {Slavin}}]{Melkov.2001}%
	\BibitemOpen
	\bibfield  {author} {\bibinfo {author} {\bibfnamefont {G.~A.}\ \bibnamefont
			{Melkov}}, \bibinfo {author} {\bibfnamefont {A.~A.}\ \bibnamefont {Serga}},
		\bibinfo {author} {\bibfnamefont {V.~S.}\ \bibnamefont {Tiberkevich}},
		\bibinfo {author} {\bibfnamefont {Y.~V.}\ \bibnamefont {Kobljanskij}},\ and\
		\bibinfo {author} {\bibfnamefont {A.~N.}\ \bibnamefont {Slavin}},\ }\bibfield
	{title} {\bibinfo {title} {{Nonadiabatic interaction of a propagating wave
				packet with localized parametric pumping}},\ }\href
	{https://doi.org/10.1103/PhysRevE.63.066607} {\bibfield  {journal} {\bibinfo
			{journal} {Phys. Rev. E}\ }\textbf {\bibinfo {volume} {63}},\ \bibinfo
		{pages} {066607} (\bibinfo {year} {2001})}\BibitemShut {NoStop}%
	\bibitem [{\citenamefont {Brächer}\ \emph {et~al.}(2014)\citenamefont
		{Brächer}, \citenamefont {Pirro}, \citenamefont {Heussner}, \citenamefont
		{Serga},\ and\ \citenamefont {Hillebrands}}]{Bracher.2014}%
	\BibitemOpen
	\bibfield  {author} {\bibinfo {author} {\bibfnamefont {T.}~\bibnamefont
			{Brächer}}, \bibinfo {author} {\bibfnamefont {P.}~\bibnamefont {Pirro}},
		\bibinfo {author} {\bibfnamefont {F.}~\bibnamefont {Heussner}}, \bibinfo
		{author} {\bibfnamefont {A.~A.}\ \bibnamefont {Serga}},\ and\ \bibinfo
		{author} {\bibfnamefont {B.}~\bibnamefont {Hillebrands}},\ }\bibfield
	{title} {\bibinfo {title} {{Localized parallel parametric generation of spin
				waves in a Ni\textsubscript{81}Fe\textsubscript{19} waveguide by spatial
				variation of the pumping field}},\ }\href {https://doi.org/10.1063/1.4867982}
	{\bibfield  {journal} {\bibinfo  {journal} {Appl. Phys. Lett.}\ }\textbf
		{\bibinfo {volume} {104}},\ \bibinfo {pages} {092418} (\bibinfo {year}
		{2014})}\BibitemShut {NoStop}%
	\bibitem [{\citenamefont {Brächer}\ \emph {et~al.}(2017)\citenamefont
		{Brächer}, \citenamefont {Pirro},\ and\ \citenamefont
		{Hillebrands}}]{Bracher.2017}%
	\BibitemOpen
	\bibfield  {author} {\bibinfo {author} {\bibfnamefont {T.}~\bibnamefont
			{Brächer}}, \bibinfo {author} {\bibfnamefont {P.}~\bibnamefont {Pirro}},\
		and\ \bibinfo {author} {\bibfnamefont {B.}~\bibnamefont {Hillebrands}},\
	}\bibfield  {title} {\bibinfo {title} {{Parallel pumping for magnon
				spintronics: Amplification and manipulation of magnon spin currents on the
				micron-scale}},\ }\href
	{https://doi.org/https://doi.org/10.1016/j.physrep.2017.07.003} {\bibfield
		{journal} {\bibinfo  {journal} {Phys. Rep.}\ }\textbf {\bibinfo {volume}
			{699}},\ \bibinfo {pages} {1} (\bibinfo {year} {2017})}\BibitemShut {NoStop}%
	\bibitem [{\citenamefont {Mohseni}\ \emph {et~al.}(2020)\citenamefont
		{Mohseni}, \citenamefont {Kewenig}, \citenamefont {Verba}, \citenamefont
		{Wang}, \citenamefont {Schneider}, \citenamefont {Heinz}, \citenamefont
		{Kohl}, \citenamefont {Dubs}, \citenamefont {Lägel}, \citenamefont {Serga},
		\citenamefont {Hillebrands}, \citenamefont {Chumak},\ and\ \citenamefont
		{Pirro}}]{Mohseni.2020}%
	\BibitemOpen
	\bibfield  {author} {\bibinfo {author} {\bibfnamefont {M.}~\bibnamefont
			{Mohseni}}, \bibinfo {author} {\bibfnamefont {M.}~\bibnamefont {Kewenig}},
		\bibinfo {author} {\bibfnamefont {R.}~\bibnamefont {Verba}}, \bibinfo
		{author} {\bibfnamefont {Q.}~\bibnamefont {Wang}}, \bibinfo {author}
		{\bibfnamefont {M.}~\bibnamefont {Schneider}}, \bibinfo {author}
		{\bibfnamefont {B.}~\bibnamefont {Heinz}}, \bibinfo {author} {\bibfnamefont
			{F.}~\bibnamefont {Kohl}}, \bibinfo {author} {\bibfnamefont {C.}~\bibnamefont
			{Dubs}}, \bibinfo {author} {\bibfnamefont {B.}~\bibnamefont {Lägel}},
		\bibinfo {author} {\bibfnamefont {A.~A.}\ \bibnamefont {Serga}}, \bibinfo
		{author} {\bibfnamefont {B.}~\bibnamefont {Hillebrands}}, \bibinfo {author}
		{\bibfnamefont {A.~V.}\ \bibnamefont {Chumak}},\ and\ \bibinfo {author}
		{\bibfnamefont {P.}~\bibnamefont {Pirro}},\ }\bibfield  {title} {\bibinfo
		{title} {{Parametric generation of propagating spin waves in ultrathin
				yttrium iron garnet waveguides}},\ }\href
	{https://doi.org/10.1002/pssr.202070022} {\bibfield  {journal} {\bibinfo
			{journal} {Phys. Status Solidi RRL}\ }\textbf {\bibinfo {volume} {14}},\
		\bibinfo {pages} {2070022} (\bibinfo {year} {2020})}\BibitemShut {NoStop}%
	\bibitem [{\citenamefont {Deka}\ \emph {et~al.}(2022)\citenamefont {Deka},
		\citenamefont {Rana}, \citenamefont {Anami}, \citenamefont {Miura},
		\citenamefont {Takahashi}, \citenamefont {Otani},\ and\ \citenamefont
		{Fukuma}}]{Deka.2022}%
	\BibitemOpen
	\bibfield  {author} {\bibinfo {author} {\bibfnamefont {A.}~\bibnamefont
			{Deka}}, \bibinfo {author} {\bibfnamefont {B.}~\bibnamefont {Rana}}, \bibinfo
		{author} {\bibfnamefont {R.}~\bibnamefont {Anami}}, \bibinfo {author}
		{\bibfnamefont {K.}~\bibnamefont {Miura}}, \bibinfo {author} {\bibfnamefont
			{H.}~\bibnamefont {Takahashi}}, \bibinfo {author} {\bibfnamefont
			{Y.C.}~\bibnamefont {Otani}},\ and\ \bibinfo {author} {\bibfnamefont
			{Y.}~\bibnamefont {Fukuma}},\ }\bibfield  {title} {\bibinfo {title}
		{{Electric field induced parametric excitation of exchange magnons in a
				CoFeB/MgO junction}},\ }\href
	{https://doi.org/10.1103/PhysRevResearch.4.023139} {\bibfield  {journal}
		{\bibinfo  {journal} {Phys. Rev. Res.}\ }\textbf {\bibinfo {volume} {4}},\
		\bibinfo {pages} {023139} (\bibinfo {year} {2022})}\BibitemShut {NoStop}%
	\bibitem [{\citenamefont {Heinz}\ \emph {et~al.}(2022)\citenamefont {Heinz},
		\citenamefont {Mohseni}, \citenamefont {Lentfert}, \citenamefont {Verba},
		\citenamefont {Schneider}, \citenamefont {L\"agel}, \citenamefont
		{Levchenko}, \citenamefont {Br\"acher}, \citenamefont {Dubs}, \citenamefont
		{Chumak},\ and\ \citenamefont {Pirro}}]{Heinz.2022}%
	\BibitemOpen
	\bibfield  {author} {\bibinfo {author} {\bibfnamefont {B.}~\bibnamefont
			{Heinz}}, \bibinfo {author} {\bibfnamefont {M.}~\bibnamefont {Mohseni}},
		\bibinfo {author} {\bibfnamefont {A.}~\bibnamefont {Lentfert}}, \bibinfo
		{author} {\bibfnamefont {R.}~\bibnamefont {Verba}}, \bibinfo {author}
		{\bibfnamefont {M.}~\bibnamefont {Schneider}}, \bibinfo {author}
		{\bibfnamefont {B.}~\bibnamefont {L\"agel}}, \bibinfo {author} {\bibfnamefont
			{K.}~\bibnamefont {Levchenko}}, \bibinfo {author} {\bibfnamefont
			{T.}~\bibnamefont {Br\"acher}}, \bibinfo {author} {\bibfnamefont
			{C.}~\bibnamefont {Dubs}}, \bibinfo {author} {\bibfnamefont {A.~V.}\
			\bibnamefont {Chumak}},\ and\ \bibinfo {author} {\bibfnamefont
			{P.}~\bibnamefont {Pirro}},\ }\bibfield  {title} {\bibinfo {title}
		{{Parametric generation of spin waves in nanoscaled magnonic conduits}},\
	}\href {https://doi.org/10.1103/PhysRevB.105.144424} {\bibfield  {journal}
		{\bibinfo  {journal} {Phys. Rev. B}\ }\textbf {\bibinfo {volume} {105}},\
		\bibinfo {pages} {144424} (\bibinfo {year} {2022})}\BibitemShut {NoStop}%
	\bibitem [{\citenamefont {Berger}(1996)}]{Berger.1996}%
	\BibitemOpen
	\bibfield  {author} {\bibinfo {author} {\bibfnamefont {L.}~\bibnamefont
			{Berger}},\ }\bibfield  {title} {\bibinfo {title} {{Emission of spin waves by
				a magnetic multilayer traversed by a current}},\ }\href
	{https://doi.org/10.1103/PhysRevB.54.9353} {\bibfield  {journal} {\bibinfo
			{journal} {Phys. Rev. B}\ }\textbf {\bibinfo {volume} {54}},\ \bibinfo
		{pages} {9353} (\bibinfo {year} {1996})}\BibitemShut {NoStop}%
	\bibitem [{\citenamefont {Slonczewski}(1996)}]{Slonczewski.1996}%
	\BibitemOpen
	\bibfield  {author} {\bibinfo {author} {\bibfnamefont {J.}~\bibnamefont
			{Slonczewski}},\ }\bibfield  {title} {\bibinfo {title} {{Current-driven
				excitation of magnetic multilayers}},\ }\href
	{https://doi.org/https://doi.org/10.1016/0304-8853(96)00062-5} {\bibfield
		{journal} {\bibinfo  {journal} {J. Magn. Magn. Mater.}\ }\textbf {\bibinfo
			{volume} {159}},\ \bibinfo {pages} {L1} (\bibinfo {year} {1996})}\BibitemShut
	{NoStop}%
	\bibitem [{\citenamefont {Padrón-Hernández}\ \emph
		{et~al.}(2011)\citenamefont {Padrón-Hernández}, \citenamefont {Azevedo},\
		and\ \citenamefont {Rezende}}]{Hernandez.2011}%
	\BibitemOpen
	\bibfield  {author} {\bibinfo {author} {\bibfnamefont {E.}~\bibnamefont
			{Padrón-Hernández}}, \bibinfo {author} {\bibfnamefont {A.}~\bibnamefont
			{Azevedo}},\ and\ \bibinfo {author} {\bibfnamefont {S.~M.}\ \bibnamefont
			{Rezende}},\ }\bibfield  {title} {\bibinfo {title} {{Amplification of spin
				waves in yttrium iron garnet films through the spin Hall effect}},\ }\href
	{https://doi.org/10.1063/1.3660586} {\bibfield  {journal} {\bibinfo
			{journal} {Appl. Phys. Lett.}\ }\textbf {\bibinfo {volume} {99}},\ \bibinfo
		{pages} {192511} (\bibinfo {year} {2011})}\BibitemShut {NoStop}%
	\bibitem [{\citenamefont {Hamadeh}\ \emph {et~al.}(2014)\citenamefont
		{Hamadeh}, \citenamefont {d'Allivy Kelly}, \citenamefont {Hahn},
		\citenamefont {Meley}, \citenamefont {Bernard}, \citenamefont {Molpeceres},
		\citenamefont {Naletov}, \citenamefont {Viret}, \citenamefont {Anane},
		\citenamefont {Cros}, \citenamefont {Demokritov}, \citenamefont {Prieto},
		\citenamefont {Mu\~noz}, \citenamefont {de~Loubens},\ and\ \citenamefont
		{Klein}}]{Hamadeh.2014}%
	\BibitemOpen
	\bibfield  {author} {\bibinfo {author} {\bibfnamefont {A.}~\bibnamefont
			{Hamadeh}}, \bibinfo {author} {\bibfnamefont {O.}~\bibnamefont {d'Allivy
				Kelly}}, \bibinfo {author} {\bibfnamefont {C.}~\bibnamefont {Hahn}}, \bibinfo
		{author} {\bibfnamefont {H.}~\bibnamefont {Meley}}, \bibinfo {author}
		{\bibfnamefont {R.}~\bibnamefont {Bernard}}, \bibinfo {author} {\bibfnamefont
			{A.~H.}\ \bibnamefont {Molpeceres}}, \bibinfo {author} {\bibfnamefont
			{V.~V.}\ \bibnamefont {Naletov}}, \bibinfo {author} {\bibfnamefont
			{M.}~\bibnamefont {Viret}}, \bibinfo {author} {\bibfnamefont
			{A.}~\bibnamefont {Anane}}, \bibinfo {author} {\bibfnamefont
			{V.}~\bibnamefont {Cros}}, \bibinfo {author} {\bibfnamefont {S.~O.}\
			\bibnamefont {Demokritov}}, \bibinfo {author} {\bibfnamefont {J.~L.}\
			\bibnamefont {Prieto}}, \bibinfo {author} {\bibfnamefont {M.}~\bibnamefont
			{Mu\~noz}}, \bibinfo {author} {\bibfnamefont {G.}~\bibnamefont
			{de~Loubens}},\ and\ \bibinfo {author} {\bibfnamefont {O.}~\bibnamefont
			{Klein}},\ }\bibfield  {title} {\bibinfo {title} {{Full vontrol of the
				spin-wave damping in a magnetic insulator using spin-orbit torque}},\ }\href
	{https://doi.org/10.1103/PhysRevLett.113.197203} {\bibfield  {journal}
		{\bibinfo  {journal} {Phys. Rev. Lett.}\ }\textbf {\bibinfo {volume} {113}},\
		\bibinfo {pages} {197203} (\bibinfo {year} {2014})}\BibitemShut {NoStop}%
	\bibitem [{\citenamefont {Evelt}\ \emph {et~al.}(2016)\citenamefont {Evelt},
		\citenamefont {Demidov}, \citenamefont {Bessonov}, \citenamefont
		{Demokritov}, \citenamefont {Prieto}, \citenamefont {Muñoz}, \citenamefont
		{Ben~Youssef}, \citenamefont {Naletov}, \citenamefont {de~Loubens},
		\citenamefont {Klein}, \citenamefont {Collet}, \citenamefont
		{Garcia-Hernandez}, \citenamefont {Bortolotti}, \citenamefont {Cros},\ and\
		\citenamefont {Anane}}]{Evelt.2016}%
	\BibitemOpen
	\bibfield  {author} {\bibinfo {author} {\bibfnamefont {M.}~\bibnamefont
			{Evelt}}, \bibinfo {author} {\bibfnamefont {V.~E.}\ \bibnamefont {Demidov}},
		\bibinfo {author} {\bibfnamefont {V.}~\bibnamefont {Bessonov}}, \bibinfo
		{author} {\bibfnamefont {S.~O.}\ \bibnamefont {Demokritov}}, \bibinfo
		{author} {\bibfnamefont {J.~L.}\ \bibnamefont {Prieto}}, \bibinfo {author}
		{\bibfnamefont {M.}~\bibnamefont {Muñoz}}, \bibinfo {author} {\bibfnamefont
			{J.}~\bibnamefont {Ben~Youssef}}, \bibinfo {author} {\bibfnamefont {V.~V.}\
			\bibnamefont {Naletov}}, \bibinfo {author} {\bibfnamefont {G.}~\bibnamefont
			{de~Loubens}}, \bibinfo {author} {\bibfnamefont {O.}~\bibnamefont {Klein}},
		\bibinfo {author} {\bibfnamefont {M.}~\bibnamefont {Collet}}, \bibinfo
		{author} {\bibfnamefont {K.}~\bibnamefont {Garcia-Hernandez}}, \bibinfo
		{author} {\bibfnamefont {P.}~\bibnamefont {Bortolotti}}, \bibinfo {author}
		{\bibfnamefont {V.}~\bibnamefont {Cros}},\ and\ \bibinfo {author}
		{\bibfnamefont {A.}~\bibnamefont {Anane}},\ }\bibfield  {title} {\bibinfo
		{title} {{High-efficiency control of spin-wave propagation in ultra-thin
				yttrium iron garnet by the spin-orbit torque}},\ }\href
	{https://doi.org/10.1063/1.4948252} {\bibfield  {journal} {\bibinfo
			{journal} {Appl. Phys. Lett.}\ }\textbf {\bibinfo {volume} {108}},\ \bibinfo
		{pages} {172406} (\bibinfo {year} {2016})}\BibitemShut {NoStop}%
	\bibitem [{\citenamefont {Uchida}\ \emph {et~al.}(2008)\citenamefont {Uchida},
		\citenamefont {Takahashi}, \citenamefont {Harii}, \citenamefont {Ieda},
		\citenamefont {Koshibae}, \citenamefont {Ando}, \citenamefont {Maekawa},\
		and\ \citenamefont {Saitoh}}]{Uchida.2008}%
	\BibitemOpen
	\bibfield  {author} {\bibinfo {author} {\bibfnamefont {K.}~\bibnamefont
			{Uchida}}, \bibinfo {author} {\bibfnamefont {S.}~\bibnamefont {Takahashi}},
		\bibinfo {author} {\bibfnamefont {K.}~\bibnamefont {Harii}}, \bibinfo
		{author} {\bibfnamefont {J.}~\bibnamefont {Ieda}}, \bibinfo {author}
		{\bibfnamefont {K.}~\bibnamefont {Koshibae}}, \bibinfo {author}
		{\bibfnamefont {K.}~\bibnamefont {Ando}}, \bibinfo {author} {\bibfnamefont
			{S.}~\bibnamefont {Maekawa}},\ and\ \bibinfo {author} {\bibfnamefont
			{E.}~\bibnamefont {Saitoh}},\ }\bibfield  {title} {\bibinfo {title}
		{{Observation of the spin Seebeck effect}},\ }\href
	{https://doi.org/https://doi.org/10.1038/nature07321} {\bibfield  {journal}
		{\bibinfo  {journal} {Nature}\ }\textbf {\bibinfo {volume} {455}},\ \bibinfo
		{pages} {778} (\bibinfo {year} {2008})}\BibitemShut {NoStop}%
	\bibitem [{\citenamefont {Padrón-Hernández}\ \emph
		{et~al.}(2012)\citenamefont {Padrón-Hernández}, \citenamefont {Azevedo},\
		and\ \citenamefont {Rezende}}]{Hernandez.2012}%
	\BibitemOpen
	\bibfield  {author} {\bibinfo {author} {\bibfnamefont {E.}~\bibnamefont
			{Padrón-Hernández}}, \bibinfo {author} {\bibfnamefont {A.}~\bibnamefont
			{Azevedo}},\ and\ \bibinfo {author} {\bibfnamefont {S.~M.}\ \bibnamefont
			{Rezende}},\ }\bibfield  {title} {\bibinfo {title} {{Amplification of spin
				waves by the spin Seebeck effect}},\ }\href
	{https://doi.org/10.1063/1.3673419} {\bibfield  {journal} {\bibinfo
			{journal} {J. Appl. Phys.}\ }\textbf {\bibinfo {volume} {111}},\ \bibinfo
		{pages} {07D504} (\bibinfo {year} {2012})}\BibitemShut {NoStop}%
	\bibitem [{\citenamefont {Merbouche}\ \emph {et~al.}(2023)\citenamefont
		{Merbouche}, \citenamefont {Divinskiy}, \citenamefont {Gouéré},
		\citenamefont {Lebrun}, \citenamefont {El-Kanj}, \citenamefont {Cros},
		\citenamefont {Bortolotti}, \citenamefont {Anane}, \citenamefont
		{Demokritov},\ and\ \citenamefont {Demidov}}]{Merbouche.2023}%
	\BibitemOpen
	\bibfield  {author} {\bibinfo {author} {\bibfnamefont {H.}~\bibnamefont
			{Merbouche}}, \bibinfo {author} {\bibfnamefont {B.}~\bibnamefont
			{Divinskiy}}, \bibinfo {author} {\bibfnamefont {D.}~\bibnamefont {Gouéré}},
		\bibinfo {author} {\bibfnamefont {R.}~\bibnamefont {Lebrun}}, \bibinfo
		{author} {\bibfnamefont {A.}~\bibnamefont {El-Kanj}}, \bibinfo {author}
		{\bibfnamefont {V.}~\bibnamefont {Cros}}, \bibinfo {author} {\bibfnamefont
			{P.}~\bibnamefont {Bortolotti}}, \bibinfo {author} {\bibfnamefont
			{A.}~\bibnamefont {Anane}}, \bibinfo {author} {\bibfnamefont {S.~O.}\
			\bibnamefont {Demokritov}},\ and\ \bibinfo {author} {\bibfnamefont {V.~E.}\
			\bibnamefont {Demidov}},\ }\href@noop {} {\bibinfo {title} {{True
				amplification of spin waves in magnonic nano-waveguides}}} (\bibinfo {year}
	{2023}),\ \Eprint {https://arxiv.org/abs/2303.04695} {arXiv:2303.04695
		[cond-mat.mes-hall]} \BibitemShut {NoStop}%
	\bibitem [{\citenamefont {Schneider}\ \emph {et~al.}(2020)\citenamefont
		{Schneider}, \citenamefont {Br{\"a}cher}, \citenamefont {Breitbach},
		\citenamefont {Lauer}, \citenamefont {Pirro}, \citenamefont {Bozhko},
		\citenamefont {Musiienko-Shmarova}, \citenamefont {Heinz}, \citenamefont
		{Wang}, \citenamefont {Meyer}, \citenamefont {Heussner}, \citenamefont
		{Keller}, \citenamefont {Papaioannou}, \citenamefont {L{\"a}gel},
		\citenamefont {L{\"o}ber}, \citenamefont {Dubs}, \citenamefont {Slavin},
		\citenamefont {Tiberkevich}, \citenamefont {Serga}, \citenamefont
		{Hillebrands},\ and\ \citenamefont {Chumak}}]{RCBEC}%
	\BibitemOpen
	\bibfield  {author} {\bibinfo {author} {\bibfnamefont {M.}~\bibnamefont
			{Schneider}}, \bibinfo {author} {\bibfnamefont {T.}~\bibnamefont
			{Br{\"a}cher}}, \bibinfo {author} {\bibfnamefont {D.}~\bibnamefont
			{Breitbach}}, \bibinfo {author} {\bibfnamefont {V.}~\bibnamefont {Lauer}},
		\bibinfo {author} {\bibfnamefont {P.}~\bibnamefont {Pirro}}, \bibinfo
		{author} {\bibfnamefont {D.}~\bibnamefont {Bozhko}}, \bibinfo {author}
		{\bibfnamefont {H.~Y.}\ \bibnamefont {Musiienko-Shmarova}}, \bibinfo {author}
		{\bibfnamefont {B.}~\bibnamefont {Heinz}}, \bibinfo {author} {\bibfnamefont
			{Q.}~\bibnamefont {Wang}}, \bibinfo {author} {\bibfnamefont {T.}~\bibnamefont
			{Meyer}}, \bibinfo {author} {\bibfnamefont {F.}~\bibnamefont {Heussner}},
		\bibinfo {author} {\bibfnamefont {S.}~\bibnamefont {Keller}}, \bibinfo
		{author} {\bibfnamefont {E.~T.}\ \bibnamefont {Papaioannou}}, \bibinfo
		{author} {\bibfnamefont {B.}~\bibnamefont {L{\"a}gel}}, \bibinfo {author}
		{\bibfnamefont {T.}~\bibnamefont {L{\"o}ber}}, \bibinfo {author}
		{\bibfnamefont {C.}~\bibnamefont {Dubs}}, \bibinfo {author} {\bibfnamefont
			{A.~N.}\ \bibnamefont {Slavin}}, \bibinfo {author} {\bibfnamefont {V.~S.}\
			\bibnamefont {Tiberkevich}}, \bibinfo {author} {\bibfnamefont {A.~A.}\
			\bibnamefont {Serga}}, \bibinfo {author} {\bibfnamefont {B.}~\bibnamefont
			{Hillebrands}},\ and\ \bibinfo {author} {\bibfnamefont {A.~V.}\ \bibnamefont
			{Chumak}},\ }\bibfield  {title} {\bibinfo {title} {{Bose--Einstein
				condensation of quasiparticles by rapid cooling}},\ }\href
	{https://doi.org/https://doi.org/10.1038/s41565-020-0671-z} {\bibfield
		{journal} {\bibinfo  {journal} {Nat. Nanotechnol.}\ }\textbf {\bibinfo
			{volume} {15}},\ \bibinfo {pages} {457} (\bibinfo {year} {2020})}\BibitemShut
	{NoStop}%
	\bibitem [{\citenamefont {Schneider}\ \emph
		{et~al.}(2021{\natexlab{a}})\citenamefont {Schneider}, \citenamefont
		{Breitbach}, \citenamefont {Serha}, \citenamefont {Wang}, \citenamefont
		{Mohseni}, \citenamefont {Serga}, \citenamefont {Slavin}, \citenamefont
		{Tiberkevich}, \citenamefont {Heinz}, \citenamefont {Br\"acher},
		\citenamefont {L\"agel}, \citenamefont {Dubs}, \citenamefont {Knauer},
		\citenamefont {Dobrovolskiy}, \citenamefont {Pirro}, \citenamefont
		{Hillebrands},\ and\ \citenamefont {Chumak}}]{Schneider.2021}%
	\BibitemOpen
	\bibfield  {author} {\bibinfo {author} {\bibfnamefont {M.}~\bibnamefont
			{Schneider}}, \bibinfo {author} {\bibfnamefont {D.}~\bibnamefont
			{Breitbach}}, \bibinfo {author} {\bibfnamefont {R.~O.}\ \bibnamefont
			{Serha}}, \bibinfo {author} {\bibfnamefont {Q.}~\bibnamefont {Wang}},
		\bibinfo {author} {\bibfnamefont {M.}~\bibnamefont {Mohseni}}, \bibinfo
		{author} {\bibfnamefont {A.~A.}\ \bibnamefont {Serga}}, \bibinfo {author}
		{\bibfnamefont {A.~N.}\ \bibnamefont {Slavin}}, \bibinfo {author}
		{\bibfnamefont {V.~S.}\ \bibnamefont {Tiberkevich}}, \bibinfo {author}
		{\bibfnamefont {B.}~\bibnamefont {Heinz}}, \bibinfo {author} {\bibfnamefont
			{T.}~\bibnamefont {Br\"acher}}, \bibinfo {author} {\bibfnamefont
			{B.}~\bibnamefont {L\"agel}}, \bibinfo {author} {\bibfnamefont
			{C.}~\bibnamefont {Dubs}}, \bibinfo {author} {\bibfnamefont {S.}~\bibnamefont
			{Knauer}}, \bibinfo {author} {\bibfnamefont {O.~V.}\ \bibnamefont
			{Dobrovolskiy}}, \bibinfo {author} {\bibfnamefont {P.}~\bibnamefont {Pirro}},
		\bibinfo {author} {\bibfnamefont {B.}~\bibnamefont {Hillebrands}},\ and\
		\bibinfo {author} {\bibfnamefont {A.~V.}\ \bibnamefont {Chumak}},\ }\bibfield
	{title} {\bibinfo {title} {{Stabilization of a nonlinear magnonic bullet
				coexisting with a Bose--Einstein condensate in a rapidly cooled magnonic
				system driven by spin-orbit torque}},\ }\href
	{https://doi.org/10.1103/PhysRevB.104.L140405} {\bibfield  {journal}
		{\bibinfo  {journal} {Phys. Rev. B}\ }\textbf {\bibinfo {volume} {104}},\
		\bibinfo {pages} {L140405} (\bibinfo {year}
		{2021}{\natexlab{a}})}\BibitemShut {NoStop}%
	\bibitem [{\citenamefont {Schneider}\ \emph
		{et~al.}(2021{\natexlab{b}})\citenamefont {Schneider}, \citenamefont
		{Breitbach}, \citenamefont {Serha}, \citenamefont {Wang}, \citenamefont
		{Serga}, \citenamefont {Slavin}, \citenamefont {Tiberkevich}, \citenamefont
		{Heinz}, \citenamefont {L\"agel}, \citenamefont {Br\"acher}, \citenamefont
		{Dubs}, \citenamefont {Knauer}, \citenamefont {Dobrovolskiy}, \citenamefont
		{Pirro}, \citenamefont {Hillebrands},\ and\ \citenamefont
		{Chumak}}]{Schneider.2021b}%
	\BibitemOpen
	\bibfield  {author} {\bibinfo {author} {\bibfnamefont {M.}~\bibnamefont
			{Schneider}}, \bibinfo {author} {\bibfnamefont {D.}~\bibnamefont
			{Breitbach}}, \bibinfo {author} {\bibfnamefont {R.~O.}\ \bibnamefont
			{Serha}}, \bibinfo {author} {\bibfnamefont {Q.}~\bibnamefont {Wang}},
		\bibinfo {author} {\bibfnamefont {A.~A.}\ \bibnamefont {Serga}}, \bibinfo
		{author} {\bibfnamefont {A.~N.}\ \bibnamefont {Slavin}}, \bibinfo {author}
		{\bibfnamefont {V.~S.}\ \bibnamefont {Tiberkevich}}, \bibinfo {author}
		{\bibfnamefont {B.}~\bibnamefont {Heinz}}, \bibinfo {author} {\bibfnamefont
			{B.}~\bibnamefont {L\"agel}}, \bibinfo {author} {\bibfnamefont
			{T.}~\bibnamefont {Br\"acher}}, \bibinfo {author} {\bibfnamefont
			{C.}~\bibnamefont {Dubs}}, \bibinfo {author} {\bibfnamefont {S.}~\bibnamefont
			{Knauer}}, \bibinfo {author} {\bibfnamefont {O.~V.}\ \bibnamefont
			{Dobrovolskiy}}, \bibinfo {author} {\bibfnamefont {P.}~\bibnamefont {Pirro}},
		\bibinfo {author} {\bibfnamefont {B.}~\bibnamefont {Hillebrands}},\ and\
		\bibinfo {author} {\bibfnamefont {A.~V.}\ \bibnamefont {Chumak}},\ }\bibfield
	{title} {\bibinfo {title} {Control of the bose-einstein condensation of
			magnons by the spin hall effect},\ }\href
	{https://doi.org/10.1103/PhysRevLett.127.237203} {\bibfield  {journal}
		{\bibinfo  {journal} {Phys. Rev. Lett.}\ }\textbf {\bibinfo {volume} {127}},\
		\bibinfo {pages} {237203} (\bibinfo {year} {2021}{\natexlab{b}})}\BibitemShut
	{NoStop}%
	\bibitem [{\citenamefont {Robinson}\ \emph {et~al.}(1970)\citenamefont
		{Robinson}, \citenamefont {Vural},\ and\ \citenamefont
		{Parekh}}]{Robinson.1970}%
	\BibitemOpen
	\bibfield  {author} {\bibinfo {author} {\bibfnamefont {B.}~\bibnamefont
			{Robinson}}, \bibinfo {author} {\bibfnamefont {B.}~\bibnamefont {Vural}},\
		and\ \bibinfo {author} {\bibfnamefont {J.}~\bibnamefont {Parekh}},\
	}\bibfield  {title} {\bibinfo {title} {{Spin-wave/Carrier-wave
				interactions}},\ }\href {https://doi.org/10.1109/T-ED.1970.16958} {\bibfield
		{journal} {\bibinfo  {journal} {IEEE Transactions on Electron Devices}\
		}\textbf {\bibinfo {volume} {17}},\ \bibinfo {pages} {224} (\bibinfo {year}
		{1970})}\BibitemShut {NoStop}%
	\bibitem [{\citenamefont {Nunes}(1982)}]{Nunes.1982}%
	\BibitemOpen
	\bibfield  {author} {\bibinfo {author} {\bibfnamefont {O.}~\bibnamefont
			{Nunes}},\ }\bibfield  {title} {\bibinfo {title} {{Spin wave amplification by
				a radiation field in free-carrier magnetic semiconductors}},\ }\href
	{https://doi.org/https://doi.org/10.1016/0038-1098(82)91102-4} {\bibfield
		{journal} {\bibinfo  {journal} {Solid State Communications}\ }\textbf
		{\bibinfo {volume} {44}},\ \bibinfo {pages} {1275} (\bibinfo {year}
		{1982})}\BibitemShut {NoStop}%
	\bibitem [{\citenamefont {Souto}\ \emph {et~al.}(2001)\citenamefont {Souto},
		\citenamefont {Nunes}, \citenamefont {Agrello},\ and\ \citenamefont
		{Fonseca}}]{Souto.2001}%
	\BibitemOpen
	\bibfield  {author} {\bibinfo {author} {\bibfnamefont {E.}~\bibnamefont
			{Souto}}, \bibinfo {author} {\bibfnamefont {O.}~\bibnamefont {Nunes}},
		\bibinfo {author} {\bibfnamefont {D.}~\bibnamefont {Agrello}},\ and\ \bibinfo
		{author} {\bibfnamefont {A.}~\bibnamefont {Fonseca}},\ }\bibfield  {title}
	{\bibinfo {title} {Spin wave amplification in antiferromagnetic
			semiconductors stimulated by infrared laser field},\ }\href
	{https://doi.org/https://doi.org/10.1016/S0375-9601(01)00441-8} {\bibfield
		{journal} {\bibinfo  {journal} {Physics Letters A}\ }\textbf {\bibinfo
			{volume} {286}},\ \bibinfo {pages} {353} (\bibinfo {year}
		{2001})}\BibitemShut {NoStop}%
	\bibitem [{\citenamefont {Collet}\ \emph {et~al.}(2016)\citenamefont {Collet},
		\citenamefont {de~Milly}, \citenamefont {d'Allivy Kelly}, \citenamefont
		{Naletov}, \citenamefont {Bernard}, \citenamefont {Bortolotti}, \citenamefont
		{Ben~Youssef}, \citenamefont {Demidov}, \citenamefont {Demokritov},
		\citenamefont {Prieto}, \citenamefont {Mu{\~{n}}oz}, \citenamefont {Cros},
		\citenamefont {Anane}, \citenamefont {de~Loubens},\ and\ \citenamefont
		{Klein}}]{Collet.2016}%
	\BibitemOpen
	\bibfield  {author} {\bibinfo {author} {\bibfnamefont {M.}~\bibnamefont
			{Collet}}, \bibinfo {author} {\bibfnamefont {X.}~\bibnamefont {de~Milly}},
		\bibinfo {author} {\bibfnamefont {O.}~\bibnamefont {d'Allivy Kelly}},
		\bibinfo {author} {\bibfnamefont {V.~V.}\ \bibnamefont {Naletov}}, \bibinfo
		{author} {\bibfnamefont {R.}~\bibnamefont {Bernard}}, \bibinfo {author}
		{\bibfnamefont {P.}~\bibnamefont {Bortolotti}}, \bibinfo {author}
		{\bibfnamefont {J.}~\bibnamefont {Ben~Youssef}}, \bibinfo {author}
		{\bibfnamefont {V.~E.}\ \bibnamefont {Demidov}}, \bibinfo {author}
		{\bibfnamefont {S.~O.}\ \bibnamefont {Demokritov}}, \bibinfo {author}
		{\bibfnamefont {J.~L.}\ \bibnamefont {Prieto}}, \bibinfo {author}
		{\bibfnamefont {M.}~\bibnamefont {Mu{\~{n}}oz}}, \bibinfo {author}
		{\bibfnamefont {V.}~\bibnamefont {Cros}}, \bibinfo {author} {\bibfnamefont
			{A.}~\bibnamefont {Anane}}, \bibinfo {author} {\bibfnamefont
			{G.}~\bibnamefont {de~Loubens}},\ and\ \bibinfo {author} {\bibfnamefont
			{O.}~\bibnamefont {Klein}},\ }\bibfield  {title} {\bibinfo {title}
		{{Generation of coherent spin-wave modes in yttrium iron garnet microdiscs by
				spin--orbit torque}},\ }\href {https://doi.org/10.1038/ncomms10377}
	{\bibfield  {journal} {\bibinfo  {journal} {Nature Communications}\ }\textbf
		{\bibinfo {volume} {7}},\ \bibinfo {pages} {10377} (\bibinfo {year}
		{2016})}\BibitemShut {NoStop}%
	\bibitem [{\citenamefont {Tsoi}\ \emph {et~al.}(2000)\citenamefont {Tsoi},
		\citenamefont {Jansen}, \citenamefont {Bass}, \citenamefont {Chiang},
		\citenamefont {Tsoi},\ and\ \citenamefont {Wyder}}]{Tsoi.2000}%
	\BibitemOpen
	\bibfield  {author} {\bibinfo {author} {\bibfnamefont {M.}~\bibnamefont
			{Tsoi}}, \bibinfo {author} {\bibfnamefont {A.~G.~M.}\ \bibnamefont {Jansen}},
		\bibinfo {author} {\bibfnamefont {J.}~\bibnamefont {Bass}}, \bibinfo {author}
		{\bibfnamefont {W.-C.}\ \bibnamefont {Chiang}}, \bibinfo {author}
		{\bibfnamefont {V.}~\bibnamefont {Tsoi}},\ and\ \bibinfo {author}
		{\bibfnamefont {P.}~\bibnamefont {Wyder}},\ }\bibfield  {title} {\bibinfo
		{title} {{Generation and detection of phase-coherent current-driven magnons
				in magnetic multilayers}},\ }\href {https://doi.org/10.1038/35017512}
	{\bibfield  {journal} {\bibinfo  {journal} {Nature}\ }\textbf {\bibinfo
			{volume} {406}},\ \bibinfo {pages} {46} (\bibinfo {year} {2000})}\BibitemShut
	{NoStop}%
	\bibitem [{\citenamefont {Doornenbal}\ \emph {et~al.}(2019)\citenamefont
		{Doornenbal}, \citenamefont {Rold\'an-Molina}, \citenamefont {Nunez},\ and\
		\citenamefont {Duine}}]{Doornenbal.2019}%
	\BibitemOpen
	\bibfield  {author} {\bibinfo {author} {\bibfnamefont {R.~J.}\ \bibnamefont
			{Doornenbal}}, \bibinfo {author} {\bibfnamefont {A.}~\bibnamefont
			{Rold\'an-Molina}}, \bibinfo {author} {\bibfnamefont {A.~S.}\ \bibnamefont
			{Nunez}},\ and\ \bibinfo {author} {\bibfnamefont {R.~A.}\ \bibnamefont
			{Duine}},\ }\bibfield  {title} {\bibinfo {title} {{Spin-Wave Amplification
				and Lasing Driven by Inhomogeneous Spin-Transfer Torques}},\ }\href
	{https://doi.org/10.1103/PhysRevLett.122.037203} {\bibfield  {journal}
		{\bibinfo  {journal} {Phys. Rev. Lett.}\ }\textbf {\bibinfo {volume} {122}},\
		\bibinfo {pages} {037203} (\bibinfo {year} {2019})}\BibitemShut {NoStop}%
	\bibitem [{\citenamefont {Danilov}(1980)}]{Danilov.1980}%
	\BibitemOpen
	\bibfield  {author} {\bibinfo {author} {\bibfnamefont {V.~V.}\ \bibnamefont
			{Danilov}},\ }\bibfield  {title} {\bibinfo {title} {{Effects of the
				interaction between surface magnetostatic waves and the spin system of a
				paramagnetic crystal}},\ }\href {https://doi.org/10.1007/BF01033471}
	{\bibfield  {journal} {\bibinfo  {journal} {Radiophysics and Quantum
				Electronics}\ }\textbf {\bibinfo {volume} {23}},\ \bibinfo {pages} {1006}
		(\bibinfo {year} {1980})}\BibitemShut {NoStop}%
	\bibitem [{\citenamefont {Danilov}\ and\ \citenamefont
		{Nechiporuk}(2002)}]{Danilov.2002}%
	\BibitemOpen
	\bibfield  {author} {\bibinfo {author} {\bibfnamefont {V.~V.}\ \bibnamefont
			{Danilov}}\ and\ \bibinfo {author} {\bibfnamefont {A.~Y.}\ \bibnamefont
			{Nechiporuk}},\ }\bibfield  {title} {\bibinfo {title} {{Experimental
				investigation of the quantum amplification effect for magnetostatic waves in
				ferrite-paramagnet structures}},\ }\href {https://doi.org/10.1134/1.1482739}
	{\bibfield  {journal} {\bibinfo  {journal} {Technical Physics Letters}\
		}\textbf {\bibinfo {volume} {28}},\ \bibinfo {pages} {369} (\bibinfo {year}
		{2002})}\BibitemShut {NoStop}%
	\bibitem [{\citenamefont {Heinz}\ \emph {et~al.}(2020)\citenamefont {Heinz},
		\citenamefont {Br{\"a}cher}, \citenamefont {Schneider}, \citenamefont {Wang},
		\citenamefont {Lagel}, \citenamefont {Friedel}, \citenamefont {Breitbach},
		\citenamefont {Steinert}, \citenamefont {Meyer}, \citenamefont {Kewenig},
		\citenamefont {Dubs}, \citenamefont {Pirro},\ and\ \citenamefont
		{Chumak}}]{Heinz.2020}%
	\BibitemOpen
	\bibfield  {author} {\bibinfo {author} {\bibfnamefont {B.}~\bibnamefont
			{Heinz}}, \bibinfo {author} {\bibfnamefont {T.}~\bibnamefont {Br{\"a}cher}},
		\bibinfo {author} {\bibfnamefont {M.}~\bibnamefont {Schneider}}, \bibinfo
		{author} {\bibfnamefont {Q.}~\bibnamefont {Wang}}, \bibinfo {author}
		{\bibfnamefont {B.}~\bibnamefont {Lagel}}, \bibinfo {author} {\bibfnamefont
			{A.~M.}\ \bibnamefont {Friedel}}, \bibinfo {author} {\bibfnamefont
			{D.}~\bibnamefont {Breitbach}}, \bibinfo {author} {\bibfnamefont
			{S.}~\bibnamefont {Steinert}}, \bibinfo {author} {\bibfnamefont
			{T.}~\bibnamefont {Meyer}}, \bibinfo {author} {\bibfnamefont
			{M.}~\bibnamefont {Kewenig}}, \bibinfo {author} {\bibfnamefont
			{C.}~\bibnamefont {Dubs}}, \bibinfo {author} {\bibfnamefont {P.}~\bibnamefont
			{Pirro}},\ and\ \bibinfo {author} {\bibfnamefont {A.~V.}\ \bibnamefont
			{Chumak}},\ }\bibfield  {title} {\bibinfo {title} {{Propagation of spin-wave
				packets in individual nanosized yttrium iron garnet magnonic conduits}},\
	}\href {https://doi.org/10.1021/acs.nanolett.0c00657} {\bibfield  {journal}
		{\bibinfo  {journal} {Nano Lett.}\ }\textbf {\bibinfo {volume} {20}},\
		\bibinfo {pages} {4220} (\bibinfo {year} {2020})}\BibitemShut {NoStop}%
	\bibitem [{\citenamefont {Dubs}\ \emph {et~al.}(2020)\citenamefont {Dubs},
		\citenamefont {Surzhenko}, \citenamefont {Thomas}, \citenamefont {Osten},
		\citenamefont {Schneider}, \citenamefont {Lenz}, \citenamefont {Grenzer},
		\citenamefont {H\"ubner},\ and\ \citenamefont {Wendler}}]{Dubs.2020}%
	\BibitemOpen
	\bibfield  {author} {\bibinfo {author} {\bibfnamefont {C.}~\bibnamefont
			{Dubs}}, \bibinfo {author} {\bibfnamefont {O.}~\bibnamefont {Surzhenko}},
		\bibinfo {author} {\bibfnamefont {R.}~\bibnamefont {Thomas}}, \bibinfo
		{author} {\bibfnamefont {J.}~\bibnamefont {Osten}}, \bibinfo {author}
		{\bibfnamefont {T.}~\bibnamefont {Schneider}}, \bibinfo {author}
		{\bibfnamefont {K.}~\bibnamefont {Lenz}}, \bibinfo {author} {\bibfnamefont
			{J.}~\bibnamefont {Grenzer}}, \bibinfo {author} {\bibfnamefont
			{R.}~\bibnamefont {H\"ubner}},\ and\ \bibinfo {author} {\bibfnamefont
			{E.}~\bibnamefont {Wendler}},\ }\bibfield  {title} {\bibinfo {title} {{Low
				damping and microstructural perfection of sub-40nm-thin yttrium iron garnet
				films grown by liquid phase epitaxy}},\ }\href
	{https://doi.org/10.1103/PhysRevMaterials.4.024416} {\bibfield  {journal}
		{\bibinfo  {journal} {Phys. Rev. Mater.}\ }\textbf {\bibinfo {volume} {4}},\
		\bibinfo {pages} {024416} (\bibinfo {year} {2020})}\BibitemShut {NoStop}%
	\bibitem [{Sup()}]{Supplemental}%
	\BibitemOpen
	\href@noop {} {\bibinfo {title} {{See Supplemental Material for
				additional information on the spin-wave dispersion relation and group
				velocity in the YIG waveguide, the temperatures reached during Joule heating
				and the respective influence on the spin-wave propagation, the frequency
				spectrum of the amplified spin-wave packet, the influence of the RF
				excitation on the amplification factor and detailed information of our
				theoretical study on the non-equilibrium dynamics of the rapid cooling
				process.}}}\BibitemShut {Stop}%
	\bibitem [{\citenamefont {Mosendz}\ \emph {et~al.}(2010)\citenamefont
		{Mosendz}, \citenamefont {Pearson}, \citenamefont {Fradin}, \citenamefont
		{Bader},\ and\ \citenamefont {Hoffmann}}]{Mosendz.2010}%
	\BibitemOpen
	\bibfield  {author} {\bibinfo {author} {\bibfnamefont {O.}~\bibnamefont
			{Mosendz}}, \bibinfo {author} {\bibfnamefont {J.~E.}\ \bibnamefont
			{Pearson}}, \bibinfo {author} {\bibfnamefont {F.~Y.}\ \bibnamefont {Fradin}},
		\bibinfo {author} {\bibfnamefont {S.~D.}\ \bibnamefont {Bader}},\ and\
		\bibinfo {author} {\bibfnamefont {A.}~\bibnamefont {Hoffmann}},\ }\bibfield
	{title} {\bibinfo {title} {{Suppression of spin-pumping by a MgO
				tunnel-barrier}},\ }\href {https://doi.org/10.1063/1.3280378} {\bibfield
		{journal} {\bibinfo  {journal} {Appl. Phys. Lett.}\ }\textbf {\bibinfo
			{volume} {96}},\ \bibinfo {pages} {022502} (\bibinfo {year}
		{2010})}\BibitemShut {NoStop}%
	\bibitem [{\citenamefont {Ruiz-Calaforra}\ \emph {et~al.}(2015)\citenamefont
		{Ruiz-Calaforra}, \citenamefont {Brächer}, \citenamefont {Lauer},
		\citenamefont {Pirro}, \citenamefont {Heinz}, \citenamefont {Geilen},
		\citenamefont {Chumak}, \citenamefont {Conca}, \citenamefont {Leven},\ and\
		\citenamefont {Hillebrands}}]{Ruiz-Calaforra.2015}%
	\BibitemOpen
	\bibfield  {author} {\bibinfo {author} {\bibfnamefont {A.}~\bibnamefont
			{Ruiz-Calaforra}}, \bibinfo {author} {\bibfnamefont {T.}~\bibnamefont
			{Brächer}}, \bibinfo {author} {\bibfnamefont {V.}~\bibnamefont {Lauer}},
		\bibinfo {author} {\bibfnamefont {P.}~\bibnamefont {Pirro}}, \bibinfo
		{author} {\bibfnamefont {B.}~\bibnamefont {Heinz}}, \bibinfo {author}
		{\bibfnamefont {M.}~\bibnamefont {Geilen}}, \bibinfo {author} {\bibfnamefont
			{A.~V.}\ \bibnamefont {Chumak}}, \bibinfo {author} {\bibfnamefont
			{A.}~\bibnamefont {Conca}}, \bibinfo {author} {\bibfnamefont
			{B.}~\bibnamefont {Leven}},\ and\ \bibinfo {author} {\bibfnamefont
			{B.}~\bibnamefont {Hillebrands}},\ }\bibfield  {title} {\bibinfo {title}
		{{The role of the non-magnetic material in spin pumping and magnetization
				dynamics in NiFe and CoFeB multilayer systems}},\ }\href
	{https://doi.org/10.1063/1.4918909} {\bibfield  {journal} {\bibinfo
			{journal} {J. Appl. Phys.}\ }\textbf {\bibinfo {volume} {117}},\ \bibinfo
		{pages} {163901} (\bibinfo {year} {2015})}\BibitemShut {NoStop}%
	\bibitem [{\citenamefont {Wang}\ \emph {et~al.}(2019)\citenamefont {Wang},
		\citenamefont {Heinz}, \citenamefont {Verba}, \citenamefont {Kewenig},
		\citenamefont {Pirro}, \citenamefont {Schneider}, \citenamefont {Meyer},
		\citenamefont {L\"agel}, \citenamefont {Dubs}, \citenamefont {Br\"acher},\
		and\ \citenamefont {Chumak}}]{QWang.2019}%
	\BibitemOpen
	\bibfield  {author} {\bibinfo {author} {\bibfnamefont {Q.}~\bibnamefont
			{Wang}}, \bibinfo {author} {\bibfnamefont {B.}~\bibnamefont {Heinz}},
		\bibinfo {author} {\bibfnamefont {R.}~\bibnamefont {Verba}}, \bibinfo
		{author} {\bibfnamefont {M.}~\bibnamefont {Kewenig}}, \bibinfo {author}
		{\bibfnamefont {P.}~\bibnamefont {Pirro}}, \bibinfo {author} {\bibfnamefont
			{M.}~\bibnamefont {Schneider}}, \bibinfo {author} {\bibfnamefont
			{T.}~\bibnamefont {Meyer}}, \bibinfo {author} {\bibfnamefont
			{B.}~\bibnamefont {L\"agel}}, \bibinfo {author} {\bibfnamefont
			{C.}~\bibnamefont {Dubs}}, \bibinfo {author} {\bibfnamefont {T.}~\bibnamefont
			{Br\"acher}},\ and\ \bibinfo {author} {\bibfnamefont {A.~V.}\ \bibnamefont
			{Chumak}},\ }\bibfield  {title} {\bibinfo {title} {{Spin pinning and
				spin-wave dispersion in nanoscopic ferromagnetic waveguides}},\ }\href
	{https://doi.org/10.1103/PhysRevLett.122.247202} {\bibfield  {journal}
		{\bibinfo  {journal} {Phys. Rev. Lett.}\ }\textbf {\bibinfo {volume} {122}},\
		\bibinfo {pages} {247202} (\bibinfo {year} {2019})}\BibitemShut {NoStop}%
	\bibitem [{\citenamefont {{Cherepanov, Vladimir and Kolokolov, Igor and L'vov,
				Victor}}(1993)}]{SAGA}%
	\BibitemOpen
	\bibfield  {author} {\bibinfo {author} {\bibnamefont {{Cherepanov, Vladimir
					and Kolokolov, Igor and L'vov, Victor}}},\ }\bibfield  {title} {\bibinfo
		{title} {The saga of {YIG}: Spectra, thermodynamics, interaction and
			relaxation of magnons in a complex magnet},\ }\href
	{https://doi.org/10.1016/0370-1573(93)90107-O} {\bibfield  {journal}
		{\bibinfo  {journal} {Physics Reports}\ }\textbf {\bibinfo {volume} {229}},\
		\bibinfo {pages} {81} (\bibinfo {year} {1993})}\BibitemShut {NoStop}%
	\bibitem [{\citenamefont {Lvov}(1994)}]{Lvov.1994}%
	\BibitemOpen
	\bibfield  {author} {\bibinfo {author} {\bibfnamefont {V.~S.}\ \bibnamefont
			{Lvov}},\ }\href@noop {} {\emph {\bibinfo {title} {{Wave Turbulence Under
					Parametric Excitation}}}},\ \bibinfo {edition} {1st}\ ed.,\ Springer Series
	in Nonlinear Dynamics\ (\bibinfo  {publisher} {{Springer-Verlag}},\ \bibinfo
	{year} {1994})\BibitemShut {NoStop}%
\end{thebibliography}
\end{document}